\documentclass[aps,prx,preprint,superscriptaddress]{revtex4-2}

\usepackage{graphicx}
\usepackage{float}
\usepackage{bm}
\usepackage[export]{adjustbox}
\usepackage{accents}
\usepackage{color}

\newcommand	{\NH}	{N_{\text{H}}}
\newcommand	{\NHH}	{N_{\text{HH}}}
\newcommand	{\EHP}	{E_{\text{HP}}}
\newcommand  {\HD}        {H_{\text{D}}}
\newcommand  {\HP}        {H_{\text{P}}}
\newcommand  {\sv}  	{\bm{s}}

\newlength{\dhatheight}
\newcommand{\doubletilde}[1]{%
    \settoheight{\dhatheight}{\ensuremath{\hat{#1}}}%
    \addtolength{\dhatheight}{-0.35ex}%
    \tilde{\vphantom{\rule{1pt}{\dhatheight}}%
    \smash{\tilde{#1}}}}

\begin{document}



\title{Using quantum annealing to design lattice proteins}



\author{Anders Irb\"ack}
\email[]{anders.irback@cec.lu.se}
\affiliation{Computational Biology \& Biological Physics, Centre for Environmental and Climate Science (CEC), Lund University, 223 62 Lund, Sweden}


\author{Lucas Knuthson}
\affiliation{Computational Biology \& Biological Physics, Centre for Environmental and Climate Science (CEC), Lund University, 223 62 Lund, Sweden}

\author{Sandipan Mohanty}
\affiliation{Institute for Advanced Simulation, J\"ulich
Supercomputing Centre, Forschungszentrum J\"ulich, D-52425 J\"ulich, Germany}

\author{Carsten Peterson}
\affiliation{Computational Biology \& Biological Physics, Centre for Environmental and Climate Science (CEC), Lund University, 223 62 Lund, Sweden}


\date{\today}

\begin{abstract}
Quantum annealing has shown promise for finding solutions to difficult
optimization problems, including protein folding.
Recently, we used the D-Wave Advantage quantum annealer to explore 
the folding problem in a coarse-grained lattice model, the HP model, in which
amino acids are classified into two broad groups:
hydrophobic (H) and polar (P). Using a set of
22 HP sequences with up to 64 amino acids, we demonstrated the fast and consistent
identification of the correct HP model ground states using the D-Wave
hybrid quantum-classical solver.
An equally relevant biophysical challenge, called the protein design problem,
is the inverse of the above, where
the task is to predict protein sequences that fold to a given structure.
Here, we approach the design problem by a two-step procedure, implemented and executed on a D-Wave machine.
In the first step, we perform a pure sequence-space search by varying the type of amino acid at each
sequence position, and seek sequences which minimize the HP-model energy of the target structure.
After mapping this task onto an Ising spin glass representation, we employ a hybrid
quantum-classical solver to deliver energy-optimal sequences for structures
with 30--64 amino acids, with a 100\,\% success rate.
In the second step, we filter the optimized sequences from the first step
according to their ability to fold to the intended structure.
In addition, we try solving the sequence optimization problem using only the quantum
processing unit (QPU), which confines us to sizes $\le$\,20, due to exponentially decreasing
success rates.
To shed light on the pure QPU results, we investigate the effects of control errors caused by an
imperfect implementation of the intended Hamiltonian on the QPU, by numerically analyzing the
Schr\"odinger equation. We find that the simulated success
rates in the presence of control noise semi-quantitatively reproduce the modest pure QPU results
for larger chains.
\end{abstract}


\maketitle


\newpage

\section{Introduction\label{sec:intro}}

Quantum annealing (QA)~\cite{Kadowaki:98, Brooke:99,Boixo:14,Fahri:01} is a promising method 
for finding good solutions to difficult optimization problems.
In this method, one aims to find the solution to an optimization problem by encoding
it into the ground state of a spin Hamiltonian.
The approach of mapping optimization problems to spin systems is not new. It was used already in the
1980s in the context
of neural networks~\cite{Hopfield:85,Peterson:88}.
By exploiting quantum fluctuations and quantum tunneling, QA offers a potentially
much faster method for minimizing spin system energies.
Technological advances like the D-Wave Advantage
quantum annealer, with over 5,000 quantum bits (qubits) and
an average connectivity of 15~\cite{McGeoch:20}, permit exploration
of the QA approach for a wide range of scientifically interesting problems, as illustrated by
recent polymer and spin glass studies~\cite{Micheletti:23,King:23}. While most of the problems 
studied concern optimization, attempts are also being made to use quantum annealers for 
sampling~\cite{Vuffray:22,Nelson:22,Sandt:23}.   

We have recently employed this machine for protein folding using the lattice-based HP (hydrophobic/polar)
model~\cite{Lau:89} as a testbed~\cite{Irback:22}.
Earlier attempts to use quantum computing for similar folding problems were based upon chain-growth,
or turn-based, algorithms~\cite{Perdomo-Ortiz:12,Robert:21}, with which non-local interactions like chain self-avoidance are challenging
to implement unless the chain length $N$ is very short ($N \lesssim 10$). In Ref.~\cite{Irback:22}, we developed
a scalable field-like representation, with qubits at all lattice sites, which made it possible to tackle chain lengths up to $N=64$
using the D-Wave hybrid quantum-classical solver.  To ensure proper chain configurations, the approach requires
penalty terms with Lagrange parameters, but is robust with respect to the choice of these parameters.

Another equally relevant problem from the bio sector is the inverse folding problem~\cite{Kuhlman:03,Bhardway:16,Yang:19},
known as protein design, where one seeks to identify \textit{a priori} unknown sequences which fold into a given structure.
Since the structure of a protein is crucial for its function, this problem is highly relevant for drug design. 
It represents a computational challenge, as it involves the exploration of both sequence and structure spaces.

Here, we tackle the design problem for lattice proteins using QA.
We use the HP model as a testbed, for which exact results are available for chain lengths
$N\leq30$~\cite{Irback:02,Holzgrafe:11}. As is commonly done,
we split the design problem into two steps. First, we generate candidate sequences that
minimize the energy in the target structure, which we refer to as sequence optimization.
Second, we determine whether or not the optimized sequences actually fold into
the target structure by either folding the sequence using our aforementioned folding method,
or by checking in the databank of all solutions~\cite{Irback:02,Holzgrafe:11}. Prior work
proposed both QA~\cite{Mulligan:20} and gate-based \cite{Khatami:23} schemes for the first of these subproblems,
sequence optimization. Here, we present and test a complete QA approach to lattice
protein design, comprising both of the above steps.

We find that the hybrid D-Wave annealer can efficiently handle
both steps, and thus provides a fast and robust approach to the HP design problem.
By contrast, relying entirely on the quantum processing unit (QPU)
yields modest performance for larger chains. In order to understand this limitation, we develop and perform
time-dependent Schrödinger equation simulations for the sequence optimization problem,
using classical high-performance computing clusters. One potential cause of
the decrease in success rate with problem size is control errors in the Hamiltonian, which may alter the ground state
and thereby lead to incorrect solutions. Indeed, when adding control noise with its strength guided by hardware data,
we obtain results that qualitatively reproduce the modest D-Wave pure QPU results.

The approach throughout this work has been evaluated using the D-Wave Advantage as the latter is,
 currently, the only available quantum annealer with an adequate qubit count. However, our methods should be valid for any quantum annealer.


\section{Methods}

In protein design, one seeks a sequence $\sv=(s_1,\ldots,s_N)$ that folds into a given target structure, $C_t$.
In general, multiple such sequences can exist, and any such sequence constitutes a valid solution.
The probability of finding the chain with sequence $\sv$ in the state $C_t$ can be written as
\begin{equation}
P_\beta(\sv)=e^{-\beta E(C_t,\sv)} \Big/ \sum_C e^{-\beta E(C,\sv)},
\label{eq:P}
\end{equation}
where $E(C_t,\sv)$ is the energy of the sequence $\sv$ in state $C_t$ (see Sec.~\ref{sec:methods_qubo}),
$\beta$ is inverse temperature and the sum runs over all possible structures $C$.
The design problem therefore translates to finding sequences near the maximum of $P_\beta(\sv)$.
Methods for this task have been developed~\cite{Irback:99,Aina:17}.
However, maximizing $P_\beta(\sv)$ involves a generally time-consuming search in both sequence
and structure spaces. Therefore, a common approach is to first minimize the
energy in the target structure over $\sv$, $E(C_t,\sv)$, followed by a filtering step to reject candidate
sequences which have a higher probability for a different structure $C_u$. Running folding
computations, which determine the most probable structure for a given sequence,
is sufficient for this purpose.
In this work, we address the design problem by this two-step procedure
rather than directly maximizing $P_\beta(\sv)$.

\subsection{HP lattice proteins\label{sec:methods_HP}}

We consider the minimal 2D lattice-based HP model for protein folding~\cite{Lau:89}, in which
the protein is represented by a self-avoiding chain of $N$ hydrophobic (H) or polar (P) beads
that interact through a pairwise contact potential. A contact between two beads is said to occur
if they are nearest neighbors on the lattice but not along the chain. The energy function can be written
as $\EHP=-\NHH$, where $\NHH$ is the number of HH contacts~\cite{Lau:89}. This definition
renders the formation of a hydrophobic core energetically favorable. The ground state may be degenerate or unique. 
For a 2D square lattice, it is known from exhaustive enumerations that
about 2\% of all HP sequences with $N\le30$ have a unique ground state~\cite{Irback:02,Holzgrafe:11}.
The availability of exact results for all sequences with $N\le 30$ makes the 2D HP model
a useful testbed for novel computational approaches.

Despite their simplicity, coarse-grained HP models are still relevant for the qualitative
insights they provide into computationally challenging problems,
such as protein folding and design
(explored here), liquid-liquid phase separation of intrinsically
disordered proteins~\cite{Nilsson:20,Statt:20}, and protein evolution 
modeling~\cite{Bornberg-Bauer:99,Aguirre:18}.

\subsection{HP sequence optimization in QUBO form\label{sec:methods_qubo}}

Given a target structure, $C_t$, we wish to minimize the energy $\EHP(C_t,\sv)$ over sequence, $\sv$,
using a D-Wave quantum annealer. To this end, the problem must be recast in QUBO form, or,
equivalently, into an Ising spin glass format. Furthermore, an auxiliary energy term needs to be included,
to control the total number of H beads, $\NH$, in a candidate
sequence of length $N$, since the all-H homopolymer sequence constitutes a
trivial solution for unbiased $\EHP(C_t,\sv)$ minimization.

The only information needed about the target structure in order to compute $\EHP(C_t,\sv)$ is its
connectivity matrix $w_{ij}$, which indicates whether two arbitrary beads $i$ and $j$ are in contact ($w_{ij}=1$) or not
($w_{ij}=0$). This holds for any model with pairwise contact interactions, irrespective of the dimensionality and
the size of the amino acid alphabet. When using the HP model, a suitable choice of total energy $E(\sv)$ to
minimize is given by
\begin{equation}\label{eq:E}
E(\sv)=-\sum_{1\le i<j\le N} w_{ij}s_is_j + \lambda\left(\sum_{i=1}^Ns_i-\NH\right)^2
\end{equation}
where $s_i$ describes whether bead $i$ is of type P ($s_i=0$) or H ($s_i=1$). In Eq.~(\ref{eq:E}), the
first term represents $\EHP(C_t,\sv)$, whereas the second term biases the total number of H beads
toward a preset value, $\NH$. The balance between the two terms is set by the parameter $\lambda$.
The 0,1 spins of Eq.~(\ref{eq:E}) can be easily transformed into Ising $\pm1$ spins without losing
the desired quadratic structure of $E$. This energy function has a much simpler structure than
the corresponding one for the folding problem in Ref.~\cite{Irback:22}, requiring only one Lagrange 
parameter $\lambda$ instead of the three in the folding study.

This Lagrange parameter $\lambda$ must be sufficiently large
for the generated sequences to acquire the desired composition,
as set by $\NH$. On the other hand, if $\lambda$ is too large, the energy
landscape becomes rugged. Examples of how the efficiency of the
hybrid quantum-classical solver varies with $\lambda$ can be found
in Sec.~\ref{sec:hybrid_results}. All hybrid production runs were carried out
using $\lambda=2.5$. For the pure QPU computations,
which are limited to smaller systems, it was possible to use a smaller value, set to $\lambda=1.1$
(Sec.~\ref{sec:qpu_results}).

Minimizing $E(\sv)$ in Eq.~(\ref{eq:E}) can be seen as a graph bisection 
problem. Unlike a spin glass with nearest-neighbor interactions, it is a fully connected system. 
Note also that compared to the HP folding problem~\cite{Irback:22}, 
much fewer spin variables are needed, since only the H/P identity of the beads (whose 
locations and contacts are fixed for the target structure) needs to be encoded.

For all instances studied in the present paper, it is possible to infer the minimum
$\EHP$ for a given $\NH$, by inspection of the bead-bead contacts present in the target structure
(see Appendices~\ref{sec:appA} and~\ref{sec:appB}).
Hence, it is possible to decide whether or not a proposed solution is correct, even without additional calculations.

\subsection{Hybrid quantum-classical computations\label{sec:methods_hybrid}}

As an alternative to pure QPU computation, the D-Wave Advantage system also offers access to a
hybrid quantum-classical solver~\cite{McGeoch:20b}. The hybrid approach uses classical solvers while
sending suitable subproblems as queries to the QPU, to speed up the execution and improve the solutions of challenging QUBO problems.
Given the limited connectivity (15) within the D-Wave Advantage architecture, the hybrid approach is particularly relevant.
With the hybrid solver, it is possible to tackle problem sizes much larger than with pure QPU computation.

Using the hybrid solver, we performed sequence optimization for three target structures with $N=30$, $N=50$
and $N=64$ (see Fig.~\ref{fig:hybrid} below).  For each target structure $C_t$, we searched for
minimum-$\EHP(C_t,\sv$) sequences $\sv$, for several fixed compositions, $\NH$. For each
combination of $C_t$  and $\NH$, we conducted a set of 10 runs, thus generating
a set of up to 10 optimized sequences.

In the hybrid approach, the run time needs to be chosen with some care. Indeed, in our previous HP folding study~\cite{Irback:22},
the success rate of the hybrid solver for $N > 30$\, was poor for short run times, while rapidly increasing to values close to 100\,\% once
the run time passed a system size dependent threshold. To determine run times for the sequence
optimization problem, we performed a set of preliminary runs for our largest target structure ($N=64$), using $\NH=42$.
The hybrid solver consistently returned sequences with the known minimum $\EHP$ (Appendix~\ref{sec:appA})
for run times ranging from 15\,s down to the shortest possible time of 3\,s, which is also the default
run time for the D-Wave Advantage hybrid solver.
Therefore, all the production runs were carried out using this default run time.

To test whether or not the generated sequences actually fold to the desired target structures, we need to perform folding calculations.
To this end, we also employ the D-Wave hybrid solver given its demonstrated power for the folding problem~\cite{Irback:22}.
Here, for a given optimized sequence $\sv_o$, the energy $\EHP(C,\sv_o)$ was minimized over chain structure $C$, using the methods and parameters
in Ref.~\cite{Irback:22} and a $10\times10$ grid.
Based on the findings in Ref.~\cite{Irback:22}, the run time was set to 4\,s, 120\,s and 300\,s for $N=30$, $N=50$ and
$N=64$, respectively. These run times are larger than the threshold times, above which the success rate was shown
to be high~\cite{Irback:22}.

\subsection{Pure QPU computations\label{sec:methods_QPU}}

The Pegasus topology of the D-Wave Advantage QPU connects each of its qubits to 15 others~\cite{McGeoch:20}.
Problems requiring higher connectivity have to be embedded into the
Pegasus graph. This embedding is done by forming ``chains'' of qubits which act as single qubits.
The strength of the coupling between the qubits within a chain is a tunable parameter, called the chain strength.
This parameter is typically chosen slightly larger than the minimum chain strength needed to avoid having too many 
chain breaks (see Sec.~\ref{sec:Schrodinger_results}).

D-Wave offers several so-called samplers for finding embeddings into the QPU topology and performing
the QPU computation. We used the \texttt{DWaveCliqueSampler}, designed for dense binary quadratic models~\cite{D-Wave}.    
It has the property that the chains representing logical qubits share a common length, which facilitates
the analysis in Sec.~\ref{sec:Schrodinger_results}.
With this method, the number of physical qubits was three times the number of logical qubits in almost all instances
studied. For the smallest system (with $N=10$), this ratio was two instead of three.
All the computations used a chain strength between 2.25 and 4.25 (Appendix~\ref{sec:appB}). The annealing time was set
to $t_f=2000\,\mu$s, its maximum allowed value. The number
of output reads per run (annealing cycles), which must be $<$ $10^6/(t_f/\mu$s), was set to 100.

\subsection{Time-dependent Schr\"odinger equation simulations\label{sec:methods_Schrodinger}}

As will be seen in Sec.~\ref{sec:qpu_results}, the pure QPU performance deteriorates rapidly with system size.
In an attempt to understand this phenomenon, we will perform quantum mechanical simulations
with the time-dependent Schrödinger equation.

Similarly, the pure QPU performance was also meager in the folding case \cite{Irback:22}.
With its simple form [Eq.~(\ref{eq:E})], the sequence design problem is better suited for analyzing the
shortcomings of the pure QPU performance as compared to the folding case.

We consider an $N$-qubit system governed by a time-dependent Hamiltonian
\begin{equation}
H(t)=a(t)\HD+b(t)\HP\,,
\label{eq:Ht}
\end{equation}
where $\HD$ and $\HP$ are the driver and problem Hamiltonians, respectively. On a D-Wave annealer,
these two terms take the forms
\begin{equation}
\HD=\sum_i \sigma^x_i\,,\qquad
 \HP=\sum_i h_i\sigma^z_i+\sum_{i<j}J_{ij}\sigma^z_i\sigma^z_j\,,
\label{eq:HDP}
\end{equation}
where $\sigma^x_i$ and $\sigma^z_i$ denote Pauli matrices.
For specificity, we will assume a linear annealing schedule given by $a(t)=1-t/t_f$ and
$b(t)=t/t_f$, where $t_f$ is the annealing time.
We will integrate the Schr\"odinger equation with this Hamiltonian in time $t$ (with $\hbar=1$)
using the formalism and algorithm described in Appendix~\ref{sec:appC}. The units 
for time and energy are arbitrary, but related through 
$\{\mathrm{unit\ of\ time}\}\times\{\mathrm{unit\ of\ energy}\}=\hbar$.   

In addition to the driver Hamiltonian $\HD$ in Eq.~(\ref{eq:HDP}), we will also consider the so-called
$XY$-mixer~\cite{Hen:16}, given by
\begin{equation}
\HD^{XY}=\frac{1}{2} \sum_{i<j}\big(\sigma_i^x \sigma_j^x +\sigma_i^y\sigma_j^y\big)\,,
\label{eq:XY}
\end{equation}
which is currently not available on D-Wave's annealers. Transitions generated by this mixer have
the property of leaving the constrained sum in Eq.~(\ref{eq:E}) unchanged. Hence, if the $H_D^{XY}$ mixer is used,
and the initial state is a uniform superposition of all states satisfying the constraint,
it would be possible to minimize $\EHP$ at a fixed $\NH$ without including the constraint term.

\subsection{Testbed -- HP target structures}

We seek HP sequences that fold to given target structures with 10--30, 50 and 64 beads. For $N\le 30$,
all HP sequences with unique ground states and the corresponding structures are known
from exhaustive enumerations~\cite{Irback:02,Holzgrafe:11}. This data can be used to
decide whether or not a generated sequence actually folds to the desired structure. To evaluate 50- and 64-bead sequences,
for which no such data are available, we determine minimum-energy structures by using the D-Wave hybrid solver as described
in Ref.~\cite{Irback:22}.


\section{Results\label{sec:results}}

Given a target structure ($C_t$) and a composition ($\NH$), we wish to find minimum-$\EHP$ sequences
by minimizing the energy $E$ in Eq.~(\ref{eq:E}) on a quantum annealer. In all
instances studied, the minimum $\EHP$ is known (Appendices~\ref{sec:appA} and~\ref{sec:appB}), so it is possible to decide whether
or not an obtained sequence is a correct solution to the sequence optimization problem.

A sequence that minimizes $\EHP$ in $C_t$ may, however, have the same or even
lower energy in other structures $C_u \ne C_t$. Whether or not this is the case can be checked against existing exact results if 
$N\le30$~\cite{Irback:02,Holzgrafe:11}, or by performing an energy
minimization in the structure space using the hybrid quantum-classical computations as described in Ref.~\cite{Irback:22}.

D-Wave offers solvers based entirely on quantum annealing, as well as a hybrid quantum-classical scheme.
In this section, we first try out the hybrid quantum-classical approach
 with success, even for large chains (Sec.~\ref{sec:hybrid_results}).
After that,  in Sec.~\ref{sec:qpu_results}, we examine the effectiveness of pure QPU
calculations, without the classical preprocessing involved in the hybrid approach,
using smaller target structures.
We observe a rapid  decrease in the pure QPU hit rate when increasing the
size of the structures. In Sec.~\ref{sec:Schrodinger_results}, we attempt to explain this
observation by numerically solving the
Schr\"odinger equation on classical computers.


\subsection{Hybrid quantum-classical computations\label{sec:hybrid_results}}

\begin{figure}[t]
\centering
   \includegraphics[width=16cm]{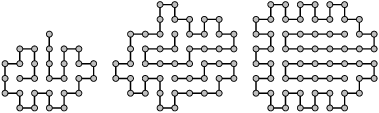}
\caption{The 30-, 50- and 64-bead target structures T$_{30}$, T$_{50}$ and T$_{64}$
used in the sequence design computations with the hybrid solver.
\label{fig:hybrid}}
\end{figure}

Using the hybrid solver, we conducted sequence design for the three target structures
shown in Fig.~\ref{fig:hybrid} with $N=30$, 50 and 64, which
will be referred to as T$_{30}$, T$_{50}$ and T$_{64}$, respectively. For each target structure, we
used a few different compositions, $\NH$ (Appendix~\ref{sec:appA}).

With no exceptions, the hybrid solver generated sequences with the known minimum
$\EHP$ (Appendix~\ref{sec:appA}) in the target structure, with a 100\,\% success rate.
 
Of note, there are sequences that minimize $\EHP$ in the target structure without folding to this structure.
For such a sequence, the target structure may be one of
multiple structures in a degenerate ground state. Alternatively, there exists at least one other
structure in which $\EHP$ is lower than it is in the target structure, so that a subsequent energy minimization
in the conformation space yields a different structure. In such cases, those sequences are not
solutions of the design problem for the target structure, and are discarded,
even when they are valid solutions for the first step of our two-step approach.
This is the price we pay for foregoing an expensive simultaneous search in sequence and structure spaces in favor of
a sequence space search as the first step. The sequences emerging from sequence space minimization
must be filtered by their ability to fold to the target structure.

Note also that our search for candidate sequences below is not exhaustive; further minimum-energy sequences  
may exist. Our goal is to find some sequence that folds into the target structure, not all such sequences.

\subsubsection{Target structure $\mathrm{T}_{30}$}

Our first target structure, T$_{30}$ (Fig.~\ref{fig:hybrid}, left panel), is known from exact
results~\cite{Holzgrafe:11} to be the unique ground state of $>$800 HP sequences.
Using the hybrid solver, we minimized $E$ in Eq.~(\ref{eq:E}), with $\lambda=2.5$,
for this structure for several compositions, $12\le\NH\le17$. For every $\NH$, 10 hybrid runs all
successfully returned sequences with the known minimum $\EHP$.

While most of the thus generated sequences had T$_{30}$ as their
unique ground state~\cite{Holzgrafe:11}, some of them (all with $\NH=12$, 13 or 16) did not.
For each of the latter, a search for possible structures with lower energy was performed, using
the hybrid solver (Sec.~\ref{sec:methods_hybrid}). No such structure was found, which suggests
that the ground states for those sequences are degenerate, and T$_{30}$ is one of the structures
having the lowest energy.

\subsubsection{Target structure $\mathrm{T}_{50}$}

The second target structure, T$_{50}$ (Fig.~\ref{fig:hybrid}, mid panel), comes from a study
of a Monte Carlo-based sequence design algorithm~\cite{Irback:99}, which actually
optimizes the target population, Eq.~(\ref{eq:P}), rather than the energy in this structure.
The best sequence found in that study contained 31 H beads~\cite{Irback:99}.

Here, we searched for sequences minimizing the energy $\EHP$ of the T$_{50}$ structure for
$\NH=29, 30$ and 31, using Eq.~(\ref{eq:E}) with $\lambda=2.5$ and the hybrid solver.  For $\NH=31$, the solution
to the $\EHP$-minimization problem is unique (Appendix~\ref{sec:appA}) and given by the sequence identified
in Ref.~\cite{Irback:99}.

As in the T$_{30}$ case, for every $\NH$, all 10 hybrid runs successfully gave
sequences with the known minimum $\EHP$ (Appendix~\ref{sec:appA}). For $\NH=29$ and 30,
where the minimum-$\EHP$ level is degenerate (Appendix~\ref{sec:appA}), the number of
distinct sequences obtained from the 10 runs were six and four, respectively.

For each of the 11 distinct optimized sequences, we subsequently minimized
$\EHP$ over structure by a set of 10 hybrid runs (Sec.~\ref{sec:methods_hybrid}).
Six of the sequences turned out to have T$_{50}$ as one of the structures at the
lowest lying $\EHP$ minimum.
For the remaining five sequences, T$_{50}$ was the only structure at the global
minimum of $\EHP$. Among them, was the sequence with $\NH=31$ from Ref.~\cite{Irback:99}. The results
obtained here support the conclusion that T$_{50}$ is the unique ground state of
the sequence found in Ref.~\cite{Irback:99}, while at the same time finding several new solutions
to the design problem for $T_{50}$.

\subsubsection{Target structure $\mathrm{T}_{64}$}

The target structure T$_{64}$ (Fig.~\ref{fig:hybrid}, right panel) has the lowest known energy for an
HP sequence with 42 H beads that has been extensively studied~\cite{Unger:93,Bastolla:98,Irback:22}.
Although not a unique energy minimum, all known structures sharing the same energy ($\EHP=-42$)
have a similar shape. The structures differ only within the entirely hydrophobic core, where the chain can
be rearranged without altering $\EHP$.

Using Eq.~(\ref{eq:E}) with $\lambda=2.5$ and the hybrid solver, we minimized $\EHP$ in the target structure T$_{64}$
for $\NH=36$, 40 and 42. As for the previous two target structures, the hybrid solver consistently
found sequences with the known minimum $\EHP$ (Appendix~\ref{sec:appA}) for all $\NH$ values.
In the $\NH=36$ and 42 cases, where multiple solutions exist (Appendix~\ref{sec:appA}), the returned sequence varied
from run to run.

As in the case of T$_{30}$ and T$_{50}$, we searched for possible structures with lower $\EHP$
using a set of 10 hybrid runs (Sec.~\ref{sec:methods_hybrid}) per optimized sequence.
For every sequence, all runs returned structures with the same $\EHP$ as the
target structure. The results thus suggest that T$_{64}$ is a
minimum-$\EHP$ structure for all these sequences. However, the minimum is not unique
for any of the optimized sequences with $\NH=40$ or 42.
For all these sequences, the core of the T$_{64}$ structure is entirely hydrophobic,
which makes it possible to rearrange the chain without changing $\EHP$.

\begin{figure}[t]
\centering
	\includegraphics[width=6cm,angle=90]{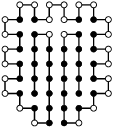}
\caption{An optimized HP sequence that appears to have the target structure
T$_{64}$ as its unique ground state. The sequence is composed of 36 H (filled) and
28 P (open) beads. For the sequence with 42 H beads studied in
Refs.~\cite{Unger:93,Bastolla:98}, this structure is one of several with minimum energy.
The degeneracy arises because, for that sequence, the chain can be rearranged in the core of
structure without altering the energy.
\label{fig:T_64_unique}}
\end{figure}

Interestingly, the situation appears to be different for one of the optimized $\NH=36$
sequences, shown in Fig.~\ref{fig:T_64_unique}, which has P beads at four core positions.
For this sequence, all 10 hybrid folding runs returned the target structure.
Changing these four beads to P in the otherwise hydrophobic core appears to
lift the degeneracy of the minimum-$\EHP$ level.

\begin{figure}[t]
\centering
	\includegraphics[width=8cm]{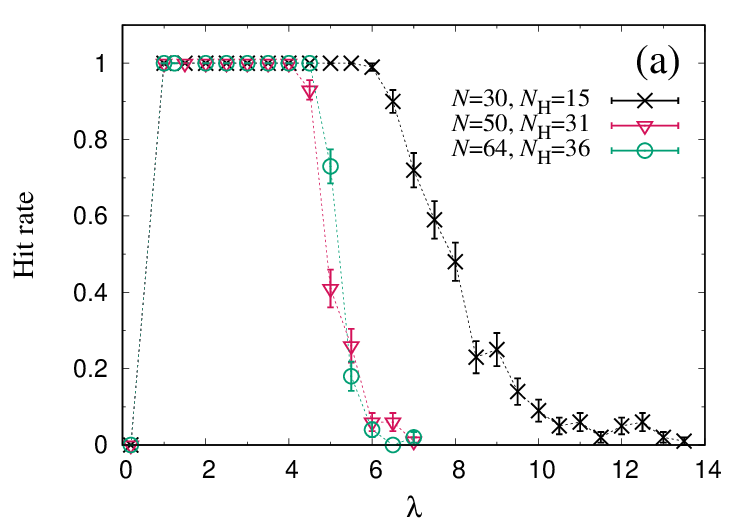}
	\includegraphics[width=8cm]{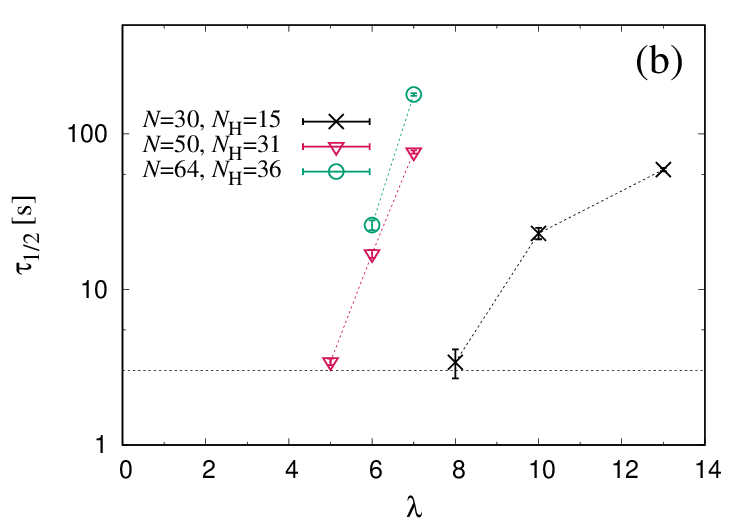}
\caption{Dependence of the hit rate on the Lagrange parameter $\lambda$ [Eq.~(\ref{eq:E})]
when solving the sequence optimization problem by hybrid quantum-classical
computations for the target structures T$_{30}$, T$_{50}$ and T$_{64}$ (Fig.~\ref{fig:hybrid})
with $\NH=15$, 31 and 36, respectively. (a) Hit rate as a function of $\lambda$ when
using the default run time, which was 3\,s for all systems. Each data point represents an average over 100 runs.
(b) The time required to attain 50\,\% hit rate, $\tau_{1/2}$, plotted on a log scale against $\lambda$.
The horizontal dotted line indicates the default run time (3\,s).
\label{fig:hit_hybrid_lambda}}
\end{figure}

The above results show that the hybrid quantum-classical method efficiently solves the
sequence optimization problem for all the systems studied. All these computations were done
with the Lagrange parameter in Eq.~(\ref{eq:E}) set to $\lambda=2.5$. To gauge the sensitivity
of the success rate to changes in $\lambda$, we conducted additional sets of hybrid runs for
the three systems (T$_{30}$, T$_{50}$ and T$_{64}$ with $\NH=15$, 31 and 36, respectively),
using the default run time. We found that $\lambda>0.25$ was necessary for any correct
solutions to be found in all three cases. The lower limit on $\lambda$ is problem-dependent but
not larger than 2.0 for any of the systems studied in this paper. As shown in [Fig.~\ref{fig:hit_hybrid_lambda}(a)],
large values for $\lambda$ also lead to performance degradation,
although, there is a wide window of $\lambda$ values having a
100\,\% hit rate,
showing that no excessive fine-tuning of $\lambda$ is required.

The low hit rates at large $\lambda$ in Fig.~\ref{fig:hit_hybrid_lambda}(a) can be improved at the
cost of increasing the run time. Figure~\ref{fig:hit_hybrid_lambda}(b) shows
the run time required to attain 50\% success rate, $\tau_{1/2}$, plotted against $\lambda$.
For the two larger problems with $N=50$ and $N=64$, respectively, at $\lambda=7$, the hit
rate is tiny when using a run time of 3\,s [Fig.~\ref{fig:hit_hybrid_lambda}(a)], but can be
improved to 50\,\% by increasing the run time to about 100\,s [Fig.~\ref{fig:hit_hybrid_lambda}(b)].
Note that $\tau_{1/2}$ grows faster with $\lambda$ for the two larger systems
than it does for the $N=30$ system. Not unexpectedly, Figs.~\ref{fig:hit_hybrid_lambda}(a,b) both
show that the $N=30$ problem is significantly easier than the other two.

To summarize, here we have designed sequences for three target structures using a two-step
procedure involving energy minimization in the sequence and structure spaces.
With this widely used approach, a pure sequence space search is performed first, where the
types of the amino acids at each position in the target structure are treated as the optimization
parameters. An optimized sequence
from the first stage is accepted as a solution to the design problem only
if a subsequent minimization in the
conformation space finds the target structure to be the global minimum for that sequence. We stress
that, for both tasks, the hybrid solver gave reliably good results, with robustness with
respect to the Lagrange parameter settings.


\subsection{Pure quantum computations\label{sec:qpu_results}}

In this section, rather than using the D-Wave hybrid solvers as in the
previous section, we explore the ability of pure QPU computations to solve the sequence
optimization problem for $10\le N\le 20$ target structures. We pick, as our target,
the most designable structure for a given $N$ denoted by $\mathrm{T}_N$,
where the designability of a structure is defined as the number of sequences sharing it
as their unique ground state. This number is known for all structures
with $N\le 30$ from exhaustive enumerations~\cite{Holzgrafe:11}. The same databank~\cite{Holzgrafe:11}
also provides exact answers to whether or not the generated
sequences actually fold to the target structures.

For each of the target structures $\mathrm{T}_{10}$--$\mathrm{T}_{20}$,
we performed pure QPU computations for one or a few choices of $\NH$ (Appendix~\ref{sec:appB}),
using the  \texttt{DWaveCliqueSampler}. As in the case of the structures in Sec.~\ref{sec:hybrid_results},
the minimum $\EHP$, given $\NH$, can be inferred from the contacts present in the target structure, and the solution
may be unique or degenerate (Appendix~\ref{sec:appB}).

\begin{figure}[t]
\centering
	\includegraphics[width=8cm]{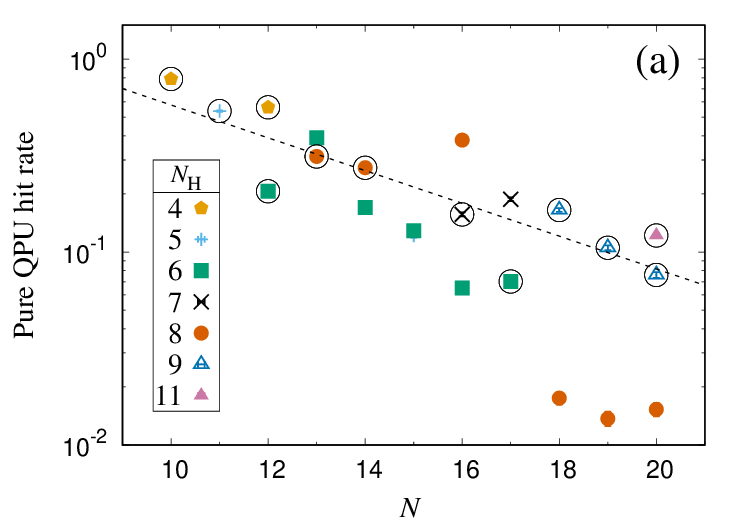}
	\includegraphics[width=8cm]{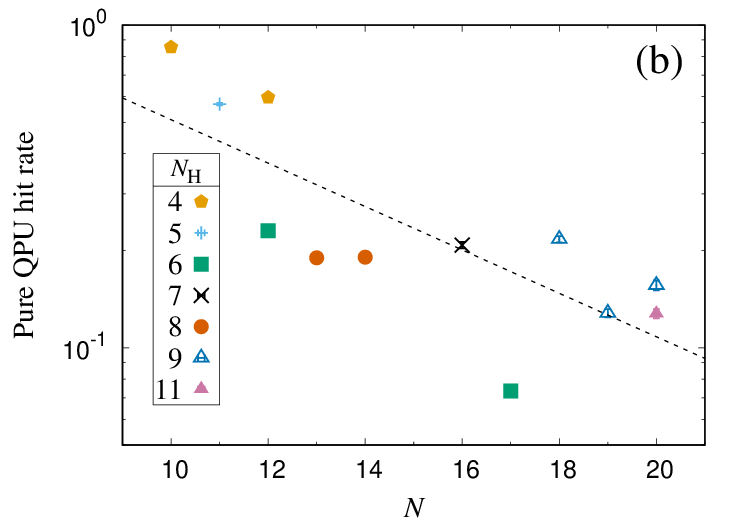}
\caption{(a) Hit rate when minimizing $E(\sv)$ in Eq.~(\ref{eq:E}) by pure QPU computation for systems with $10\le N\le 20$
and different $\NH$ (Appendix~\ref{sec:appB}). The value of $\NH$ is indicated by the plot symbol, and the Lagrange parameter was set
to $\lambda=1.1$. For each system, a set of 10,000 annealing cycles was generated using the \texttt{DWaveCliqueSampler},
and the hit rate is the fraction of these that gave a correct solution. The annealing time was set to its maximum
value, $t_f=2000$\,$\mu$s. The chain strength was chosen individually for each system among the values 2.25, 2.50,\ldots, 4.25 for best performance.
A circle around the plot symbol indicates that the system has a large energy gap, $\Delta E\ge 1.0$ (Appendix~\ref{sec:appB}).
The dashed line is a least-square fit to the encircled data points. (b) Same as panel (a), after restricting the analysis to a
filtered dataset without chain breaks (77\% of the full dataset) and removing systems with a small energy gap ($\Delta E< 1.0$).
\label{fig:hit_qpu}}
\end{figure}

The pure QPU computations recovered all possible solutions to the sequence
optimization problem for every ($\mathrm{T}_{N}, \NH$) pair. To quantify the success
 rate of the pure QPU computations, 10,000 annealing cycles
were generated for each combination of $\mathrm{T}_{N}$ and $\NH$. The fraction
of these yielding a correct solution is referred to as the hit rate as was done in Sec.~\ref{sec:hybrid_results}. For this purpose, a correct solution is a solution sequence
whose energy matches the previously known minimal $\EHP$ for the  $(\mathrm{T}_{N}, \NH)$ pair.
The hit rates obtained in the pure QPU runs can be found in Fig.~\ref{fig:hit_qpu}(a).

As noted previously, a sequence that minimizes the energy in the target structure, for a given $\NH$,
may not fold to that structure. However, for every ($\mathrm{T}_N$, $\NH$) combination
studied here, there is at least one solution to the sequence optimization problem that has
the target structure $\mathrm{T}_N$ as its unique ground state. The precise number
of such solutions for different ($\mathrm{T}_N$, $\NH$) pairs can be found in Appendix~\ref{sec:appB}.

For the generated sequences that did not have the target structure as its unique ground state,
we minimized $\EHP$ over structure using the hybrid solver
(Sec.~\ref{sec:methods_hybrid}). In these runs, for almost all the sequences, we found other structures with
the same energy as the target structure but none with lower energy, suggesting that
the target structure is part of a degenerate ground state. The only exceptions occurred
for the target structure T$_{13}$ and $\NH=6$. In this case, five of 18 minimum-$\EHP$ sequences
attained a lower energy in other structures. Nothing in our procedure precludes the existence of
lower-energy structures for a sequence obtained by minimizing the target structure energy. However, had such situations
been more common, the recipe used here (sequence space optimization followed by filtering
based on folding runs) would not be effective. Note that for $\mathrm{T}_{13}$
and $\NH=6$, the existence of lower-energy alternatives is intuitively unsurprising, considering that the target
structure has a relatively high energy for the given $\NH$ (Appendix~\ref{sec:appB}).

Although the pure QPU computations recovered all possible solutions to the sequence
optimization problem, the hit rate was strongly problem-dependent [Fig.~\ref{fig:hit_qpu}(a)].
While also depending on $\NH$, the hit rate shows a clear decreasing
trend with the problem size $N$, which limits the range of $N$ which is meaningful to study.

There are several factors that may contribute to the rapid decay of the pure QPU hit rate with problem size
as seen in Fig.~\ref{fig:hit_qpu}(a). These include (i) finite-$t_f$ effects, (ii) chain breaks, (iii) thermal noise and (iv) control errors.
Here, we briefly comment on factors (i-iii), whereas factor (iv), which relates to the implementation of the couplings $J_{ij}$ and $h_i$
in Eq.~(\ref{eq:E}), will be discussed in
Sec.~\ref{sec:Schrodinger_results} below.

\begin{figure}[t]
\centering
\begin{tabular}{llcc}
     \includegraphics[width=8cm]{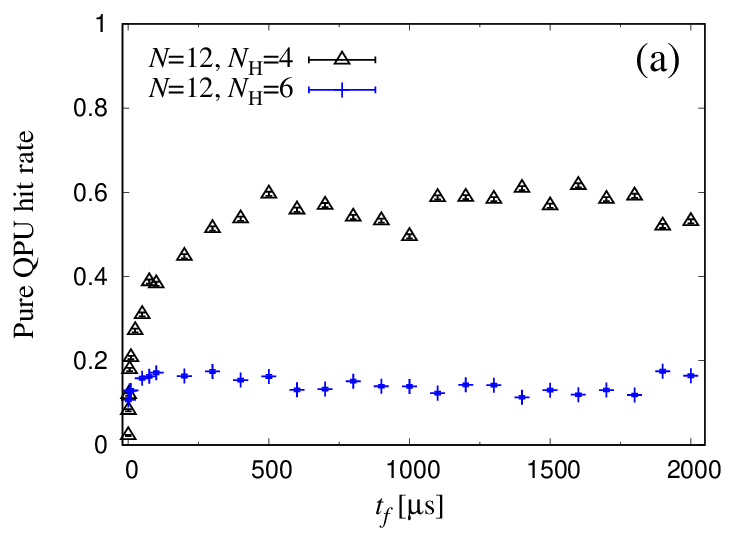}
     \vspace{4.9cm}
\end{tabular}
    \vspace{-5.5cm}
    \includegraphics[width=8cm]{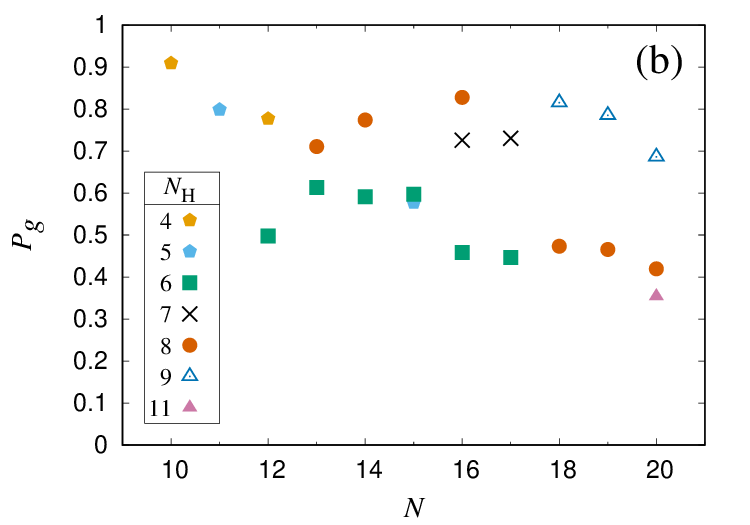}
\caption{Effects of using a finite annealing time $t_f$ in solving the sequence optimization problem
[Eq.~(\ref{eq:E}) with $\lambda=1.1$].
(a) Hit rate against $t_f$ in pure QPU computations for two of the problems
in Fig.~\ref{fig:hit_qpu} (target structure $\mathrm{T}_{12}$, $\NH= 4$ and 6).
(b) Ground-state probability at $t=t_f$, $P_g$, when numerically integrating
the Schr\"odinger equation (Sec.~\ref{sec:methods_Schrodinger}) for
systems with $10\le N\le 20$ (Appendix~\ref{sec:appB}) using a fixed finite $t_f$ ($t_f=20$\,a.u.), plotted against system size, $N$.
The value of $\NH$ is indicated by the plot symbol. The Hamiltonian is $H(t)=a(t)\HD+b(t)\HP$
[Eqs.~(\ref{eq:Ht},\ref{eq:HDP})].
\label{fig:finite_tf}}
\end{figure}

(i) \textit{Finite $t_f$.} An obvious potential source of error is the use of a finite annealing time $t_f$ ($\le$\,2000\,$\mu$s).
The $t_f$-dependence of the pure QPU hit rate is illustrated in Fig.~\ref{fig:finite_tf}(a) by data
obtained for the target structure $\mathrm{T}_{12}$ and two values of $\NH$ (4 and 6). In both cases, the hit rate
does increase with $t_f$ for small $t_f$. However, it levels off around $t_f=400$\,$\mu$s for
$\NH=4$ and already before $t_f=100$\,$\mu$s for $\NH=6$, at values well below unity.
This behavior suggests that there must be other, more important error sources than the upper limit on $t_f$.
This conclusion is further supported by results from numerically integrating the Schr\"odinger
equation (Sec.~\ref{sec:methods_Schrodinger}). Here, we computed the ground-state
probabilities, $P_g$, for the different systems at $t=t_f$, for a fixed $t_f$ [Fig.~\ref{fig:finite_tf}(a)].
With our choice of $t_f=20$\,a.u., the $P_g$ data roughly match the measured pure QPU hit rates for
small $N$ [Fig.~\ref{fig:hit_qpu}(a)] but not the rapid decay with $N$ seen in the latter case.

(ii) \textit{Chain breaks.} When embedding the problems into the Pegasus graph, the pure QPU solver creates
chains of strongly coupled physical qubits, collectively behaving as a single qubit. In computations, it may happen that such chains,
representing logical qubits, break. To check how our results are affected by such chain breaks, we recomputed the pure QPU hit rates
using data only from annealing cycles in which no chain break occurred (77\% of the full dataset).
A scatter plot comparing the original and recomputed pure QPU hit rates can be found in Appendix~\ref{sec:appD}.
In many cases, the removal of chain breaks leads to a statistically significant change of the measured hit rate,
although the overall agreement between the two datasets is quite good (Pearson correlation coefficient 0.94),

(iii) \textit{Thermal noise.}
The effects of thermal noise are likely to be more severe if the energy gap, $\Delta E$, between the ground state
and the first excited state of the problem Hamiltonian is small. In our systems, it turns out that $\Delta E$ takes on
one of three possible values, namely 0.1, 1.0 and 1.1 (Appendix~\ref{sec:appB}). In Fig.~\ref{fig:hit_qpu}(a) above,
there are three systems with markedly lower hit rates than the others, all of which have a small energy gap, $\Delta E=0.1$.
It is conceivable that the low hit rates for these systems, at least in part, is due to thermal noise. However, modeling
thermal effects is difficult without extensive details of the underlying physics of
the D-Wave processors.

In what follows, we remove from the analysis systems where thermal effects are potentially
much stronger than in the others, by focusing on systems with $\Delta E\ge 1.0$. Furthermore, we will use the
filtered dataset without chain breaks, for a cleaner comparison with results from Schr\"odinger simulations.
Redrawing Fig.~\ref{fig:hit_qpu}(a) after making these two restrictions, we obtain Fig.~\ref{fig:hit_qpu}(b), where
the pure QPU hit rate falls off  roughly exponentially with $N$, albeit still with some scatter.


\subsection{Probing the effects of control noise on the pure QPU success rate \label{sec:Schrodinger_results}}

One potentially limiting factor in the pure QPU computations is analog control errors in
the fields $h_i$ and couplers $J_{ij}$~\cite{Albach:19,Pearson:19}
of the Ising Hamiltonian $\HP$ [Eq.~(\ref{eq:HDP})],  which are referred to as integrated
control errors in D-Wave's documentation~\cite{D-Wave}. The presence of control errors,
$\delta  h_i$ and $\delta J_{ij}$, leads to a perturbed Hamiltonian
\begin{equation}
 \label{eq:ce}
	\tilde H_\mathrm{P}=\sum_{i}(h_i+\delta h_i)\sigma_i^z+\sum_{i<j}(J_{ij}+\delta J_{ij})\sigma_i^z\sigma_j^z,
\end{equation}
whose ground state may not coincide with that of the intended Hamiltonian $\HP$ [Eq.~(\ref{eq:HDP})].
Previous work showed that small errors in individual parameters collectively can cause an
exponential decay of the success rate with problem size~\cite{Albach:19,Pearson:19}.

To be able to explore the effects of control errors in our pure QPU computations,
we have to make some simplifying assumptions. First, following Refs.~\cite{Albach:19,Pearson:19},
we assume that all errors $\delta h_i$ and $\delta J_{ij}$ are statistically independent and normally distributed,
with zero mean and standard deviations $\sigma_h$ and $\sigma_J$ for all $\delta h_i$ and $\delta J_{ij}$, respectively.
Second, for computational reasons, we consider only logical qubits, thus essentially ignoring the auxiliary
qubits needed when embedding the problems into the QPU topology. However, we take the QPU embedding into
account in setting the values of $\sigma_h$
and $\sigma_J$ (see below).

In D-Wave QPU computations, all couplers and fields are rescaled to
$\hat h_i=h_i/r$ and $\hat J_{ij}=J_{ij}/r$, where $r$ is the smallest number such that all
rescaled parameters fall in given intervals, $|\hat h_i|\le h_{\max}$ and $|\hat J_{ij}|\le J_{\max}$.
In all systems studied here, the rescaling factor $r$ is set by the chain strength $J_{cs}$
(see Sec.~\ref{sec:methods_QPU}) and
given by $r=J_{cs}/J_{\max}$. In particular, this implies that the energy gap of the
rescaled problem Hamiltonian scales as $1/J_{cs}$, which should lead to a decrease
in success rate with increasing $J_{cs}$.

The strength of the control noise on D-Wave's systems has been investigated~\cite{D-Wave}.
D-Wave Support suggests using $\sigma_h=x\max|h_i|$ and $\sigma_J=x\max |J_{ij}|$ with $x=0.015$,
where the maxima are taken over all fields and couplers of the Hamiltonian, including
those associated with auxiliary qubits.
As indicated above, in our systems, the largest $|J_{ij}|$ is the chain strength $J_{cs}$.
Following the suggestion above, we therefore set $\sigma_J=xJ_{cs}$. Each logical qubit 
is represented by a chain of $k$ physical qubits (Sec.~\ref{sec:methods_QPU}), 
where $k=2$ for the $N=10$ system and $k=3$ for all other systems studied.
As the physical qubits representing a logical qubit with field $h_i$ have
fields $h_i/k$, we set $\sigma_h=\sqrt{k}\times x\max |h_i|/k$, where the square
root comes from summing over $k$ physical qubits. Summarizing, we then have
\begin{equation}
\sigma_h=\frac{x\max|h_i|}{\sqrt{k}}\quad\mathrm{and}\quad\sigma_J=xJ_{cs}\,.
\label{eq:sigma}
\end{equation}

Using Eq.~(\ref{eq:sigma}) with $k$ and $J_{cs}$ as in the pure
QPU computations and $x=0.015$, we generated 10,000 perturbed Hamiltonians $\tilde H_\mathrm{P}$
[Eq.~(\ref{eq:ce})] for each of the systems in Fig.~\ref{fig:hit_qpu}(b), and determined the fraction
of these, $P_g$, that shared ground state with the intended Hamiltonian $\HP$. For the two systems
with $(N,\NH)=(10,4)$ and $(N,\NH)=(20,11)$, we performed additional sets of pure QPU
computations to elucidate how the hit rate depends on $J_{cs}$.

\begin{figure}[t]
\centering
    \includegraphics[width=8cm]{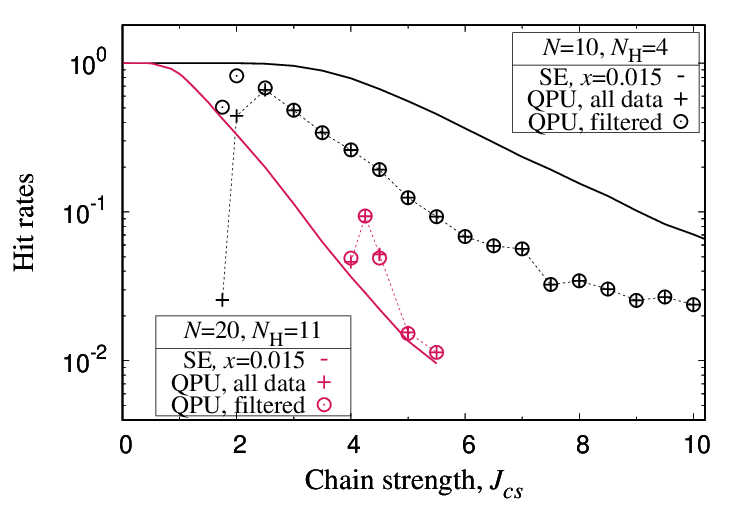}
\caption{The dependence of the hit rate on the chain strength, $J_{cs}$, in pure QPU computations
and in Schr\"odinger simulations with control noise of strength $x=0.015$, for two systems with
$(N,\NH)=(10,4)$ (black) and $(N,\NH)=(20,11)$ (red), respectively (Appendix~\ref{sec:appB}).
Full lines represent data from the Schr\"odinger simulations. Plot symbols show pure QPU hit
rates, calculated over the full dataset (plus) or the filtered dataset without chain breaks (circle).
Dotted lines are drawn to guide the eye. Data presented elsewhere in the paper for these systems
were obtained using $J_{cs}=2.25$ for $(N,\NH)=(10,4)$ and $J_{cs}=4.25$ for $(N,\NH)=(20,11)$.
\label{fig:jcs_dep}}
\end{figure}

Figure~\ref{fig:jcs_dep} shows the $J_{cs}$-dependence of the pure QPU hit rate for these two systems,
along with the simulated ground-state probability $P_g$.
Clearly, $J_{cs}$ must be sufficiently large to ensure chain stability. On the other hand, we expect
the hit rate to drop if $J_{cs}$ gets too large, due to the $1/J_{cs}$ scaling of the energy gap (see above).
The data confirm that $J_{cs}$ must be neither too small nor too large for good performance (Fig.~\ref{fig:jcs_dep}), and
therefore needs to be chosen with some care. It can also be seen that the hit rate depends only weakly on whether
the full dataset or the filtered one without chain breaks is used. As chain breaks do not occur in the
simulated systems with only logical qubits, $P_g$ does not drop at small $J_{cs}$. At large
$J_{cs}$, $P_g$ decreases at a rate similar to what is observed for the pure QPU hit rate.

\begin{figure}[t]
\centering
   \includegraphics[width=8cm]{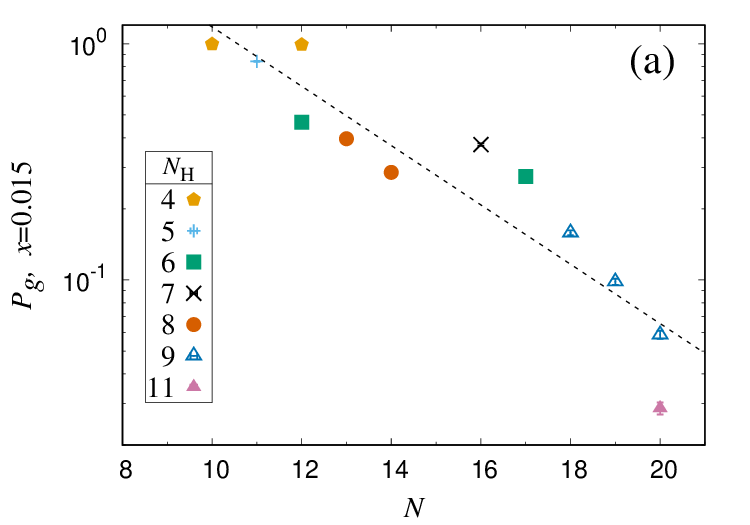}
  \includegraphics[width=8cm]{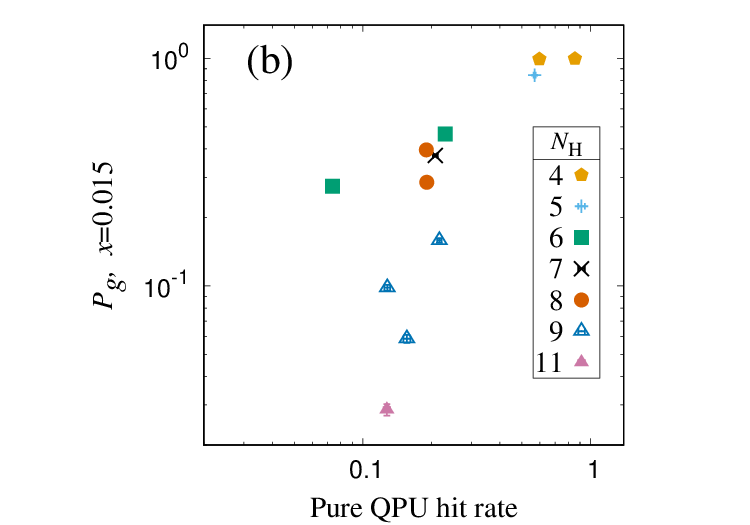}
\caption{(a) $N$-dependence of hit rates from Schr\"odinger simulations with control noise of strength $x=0.015$.
The function optimized is $E(\sv)$ in Eq.~(\ref{eq:E}) with $\lambda=1.1$ and different $\NH$, as indicated by the plot symbol.
The systems studied are the same as in Fig.~\ref{fig:hit_qpu}(b).
For each system, 10,000 perturbed Hamiltonians $\tilde H_\mathrm{P}$ [Eq.~(\ref{eq:ce})]
were generated, and the hit rate
is the fraction of these that share ground state with the noise-free Hamiltonian, $\HP$. The dashed line is a least-square fit.
(b) Scatter plot comparing hit rates from the Schr\"odinger simulations in panel (a)
with the pure QPU hit rates in Fig.~\ref{fig:hit_qpu}(b).
\label{fig:hit_noise_1.5}}
\end{figure}

Figure~\ref{fig:hit_noise_1.5}(a) shows the simulated hit rates $P_g$, for the systems studied in
Fig.~\ref{fig:hit_qpu}(b), plotted against $N$. As in the pure QPU computations [Fig.~\ref{fig:hit_qpu}(b)],
the hit rate falls off roughly exponentially with $N$. The scatter plot in Fig.~\ref{fig:hit_noise_1.5}(b)
compares simulated hit rates $P_g$ [Fig.~\ref{fig:hit_noise_1.5}(a)] with pure QPU hit rates [Fig.~\ref{fig:hit_qpu}(b)].
The pure QPU hit rate appears to decrease somewhat more slowly than $P_g$ with $N$.
Nevertheless, at a semi-quantitative level, the pure QPU hit rates agree quite well with the
$P_g$ values obtained using the suggested noise strength $x=0.015$. While refraining from 
attempts to fine-tune $x$, we note that using $x=0.003$ or $x=0.030$ leads to, respectively, 
too high or too low  $P_g$ values, compared to the QPU hit rates. 

In conclusion, the results presented here are consistent with the hypothesis that control errors are an important factor behind the strong $N$-dependence of
the pure QPU hit rates. To fully account for the measured hit rates, and in particular their $\NH$-dependence, additional factors need to be considered,
including thermal noise.

There are complementary methods, not explored here, to improve on modest pure QPU hit rates.   
One is to add a post-processing step, in which the QPU output state is subject to a local
optimization with a greedy classical algorithm~\cite{D-Wave2}, to drive approximately correct solutions to the 
desired minimum-energy level. However, in most systems studied here (Fig.~\ref{fig:hit_noise_1.5}), all first excited 
states are local minima left unchanged by this method, which suggests that it 
is of limited use for our systems. Another approach, called
``shimming'', is to refine the calibration of the Hamiltonian by using symmetries of the system~\cite{Chern:23}.  
These symmetries typically assume zero fields, and coupler values related by a sign flip ($\pm J$). These conditions 
are not met in our systems, which makes useful symmetries hard to find. 

Let us finally mention that we also investigated whether the $XY$-mixer offers an advantage in our problem, by
comparing the time evolution of the Schrödinger system when using respectively $\HD$ and $\HD^{XY}$ in
Eqs.~(\ref{eq:HDP},\ref{eq:XY}) as drivers. In the cases studied, we found that the improvement is at most marginal.


\section{Summary and Outlook}
Protein design, determining sequences corresponding to a given structure, is a highly relevant biophysical problem.
Since both sequence and structure space have to be explored, the problem is very challenging, especially for large chains.
We have approached this problem for the HP lattice protein model using quantum annealing,
by first minimizing the target structure energy to generate sequences, and then checking if the generated sequences do in fact fold to the target structure.

The approach was evaluated by using the D-Wave Advantage hybrid quantum-classical solver for three
structures with chain lengths $N=30$, $N=50$ and $N=64$.
Without exceptions, the D-Wave hybrid solver swiftly solves the sequence optimization problem,
for which the ground state energy can be deduced (Appendix~\ref{sec:appA}).
These solutions were then tested for their ability to fold to the target structure,
a problem that can also be efficiently tackled using the hybrid solver~\cite{Irback:22}. The
two-step procedure was successfully applied to all three target structures. In particular,
we identified a previously unknown sequence that appears to have the $N=64$ structure
as its unique ground state.

In addition, we tested using the D-Wave pure QPU for sequence optimization problems with $10\leq N \leq 20$,
for which solutions exist in the databank.
For all structures, sequences that had the desired structure as their unique ground state were found.
However, in line with previous results for the folding problem~\cite{Irback:22},  when using only the QPU
for the sequence optimization problem, the performance deteriorates with system size.
In order to understand this behavior, which very likely is due to hardware-induced phenomena (noise),
we solved the time-dependent Schrödinger equation numerically for different scenarios.

Two possible sources of error are inadequate annealing time and control errors in the couplers and fields
of the problem Hamiltonian. Whereas inadequate annealing time turned out not to be the problem,
our results suggest that control errors have a significant impact on the success rate.
Here, we computed the fraction of perturbed Hamiltonians sharing ground state with the original Hamiltonian
and compared with the pure QPU results.
Employing the same rescaling as on the D-Wave machine and using standard deviations of the noise supplied by D-Wave,
we found a semi-quantitative agreement with the observed pure QPU hit rates [Fig.~\ref{fig:hit_noise_1.5}(b)].
A more detailed error model should also incorporate thermal noise.
The latter would have to be based on deeper knowledge about the inner workings of the D-Wave architecture, which is,
at present, not accessible.

These noise investigations were exclusively probing the D-Wave QPU properties for the simple reason that D-Wave
is the only available annealer for realistic computations. However, our approach to elucidate the problem by comparing
with the time-dependent Schrödinger equation is of wider relevance.

Overall, it is clear that the quantum annealer naturally lends itself to protein design. Using the hybrid quantum-classical annealer,
both generating sequences, and subsequently filtering them based on their folding ability, work very well.
When using just the QPU, without leveraging D-Wave's hybrid solver, we found that generating
sequences with the lowest HP energy for the target structure becomes difficult for larger structures. The results presented
lend support to the notion that, at least in part, this difficulty originates from imperfect implementation of the problem Hamiltonian.

Replacing the $X$-driver with an $XY$-driver in the Schrödinger simulations, which has been suggested for quantum
annealing in general \cite{Hen:16}, did not improve the results for the sequence optimization problem.
This is somewhat surprising since in our case it would remove the only constraint term in Eq.~(\ref{eq:E}).

Toward more realistic protein models, one could replace the binary HP encoding spins $s_i$ by discrete multi-state spins
$S_i$ encoding both amino acid type, \textit{e.g.} in the canonical 20-letter alphabet, and the corresponding sidechain
conformations, or rotamers~\cite{Mulligan:20}. Assuming an energy with the quadratic structure
$E=\sum_{i=1}^N A(S_i)+\sum_{i<j}B(S_i,S_j)$ and a given target backbone conformation, one could then determine
the amino acid sequence and rotamers by minimizing $E$ over the $S_i$ variables. Given all possible values of all
one- and two-body terms $A(S_i)$ and $B(S_i,S_j)$, which would have to be predetermined, this minimization could in
principle be carried out using QA and a one-hot encoding of the spins $S_i$. However, checking whether or not the
optimized sequences fold to the intended and real, rather than lattice-based, backbone conformation would require
classical computing. One possibility would be to use the AlphaFold structure prediction method~\cite{Jumper:21,Varadi:21}.

\clearpage

\appendix

\section{Minimum $\EHP$ for the target structures T$_{30}$, T$_{50}$ and T$_{64}$\label{sec:appA}}

\begin{figure}[t]
\centering
   \includegraphics[width=16cm]{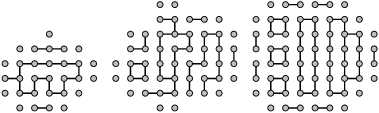}
\caption{Contact maps showing all contacts present in the 30-, 50- and 64-bead target
structures T$_{30}$, T$_{50}$ and T$_{64}$ (Fig.~\ref{fig:hybrid}). Two beads are said to be in
contact if they are nearest neighbors on the lattice but not along the chain.
\label{fig:64cmap}}
\end{figure}

Given a target structure $C_t$, we determine sequences, $\sv$, by minimizing
$\EHP(C_t,\sv)$ at different fixed compositions ($\NH$), with $\EHP$
defined as minus the number of HH contacts. In all instances studied in this paper,
the solution to this problem can be inferred from a contact map showing
all contacts present in the target structure. Figure~\ref{fig:64cmap} shows
contact maps for the three largest target structures studied:
T$_{30}$, T$_{50}$ and T$_{64}$ (Sec.~\ref{sec:hybrid_results}).

To illustrate how the minimum $\EHP$ can be found, consider the contact map
for T$_{64}$ (Fig.~\ref{fig:64cmap}, right panel) for the three $\NH$ values
used (Sec.~\ref{sec:hybrid_results}), namely 36, 40 and 42. In this structure, 10 of the beads do
not form any contact, and can therefore be ignored. The remaining beads can be
divided into 12 distinct clusters: seven two-bead clusters, three four-bead clusters
and two larger clusters with 12 and 16 beads, respectively.  The five clusters with
more than two beads contain 40 beads in total. It is easy to see that taking these 40
beads as H represents a unique $\EHP$ minimum for $\NH=40$ ($\EHP=-41$).
For $\NH=42$, the minimum $\EHP$ is obtained by adding any of the seven
two-bead clusters to the set of H beads ($\EHP=-42$). Finally, by instead removing
one of the three four-bead clusters from the $\NH=40$ solution, one finds
the minimum $\EHP$ for $\NH=36$ ($\EHP=-37$).

The T$_{30}$ and T$_{50}$ problems can be analyzed in a similar way.
In Table~\ref{tab:hybrid_problems}, we summarize
minimum $\EHP$ values and the degeneracy of the solutions for all problems
considered in our hybrid computations.

\begin{table}[h]
  \centering
  \caption{Minimum $\EHP$ and the degeneracy of the solution for all
  sequence optimization instances studied in Sec.~\ref{sec:hybrid_results},
  using the target structures T$_{30}$, T$_{50}$ and T$_{64}$ and different
  compositions, $\NH$.
  \label{tab:hybrid_problems}}
  \vspace{6pt}
  \begin{tabular}{c@{\hspace{.5cm}}c@{\hspace{.5cm}}c@{\hspace{.5cm}}c}
    \hline
    Target structure & $\NH$ & Minimum $\EHP$ & Degeneracy\\

    \hline
    T$_{30}$	& 12	& $-11$	& 14	\\
                    	& 13 	& $-12$	& 14	\\
                    	& 14	& $-14$	& 1 	\\
                    	& 15	& $-15$	& 1 	\\
                    	& 16 	& $-15$	& 18	\\
                    	& 17 	& $-16$	& 4	\\
    \hline
    T$_{50}$	& 29	& $-28$ 	& 10 	\\
                    	& 30 & $-29$ 	& 4   	\\
                    	& 31 & $-30$ 	& 1	\\
     \hline
     T$_{64}$     	& 36	& $-37$ 	& 3	\\
                    	& 40 & $-41$ 	& 1	\\
                    	& 42 & $-42$	& 7	\\
    \hline
  \end{tabular}
\end{table}

\section{Minimum $\EHP$ for the target structures with $N\le20$\label{sec:appB}}

\begin{figure}[t]
\centering
   \includegraphics[width=12cm]{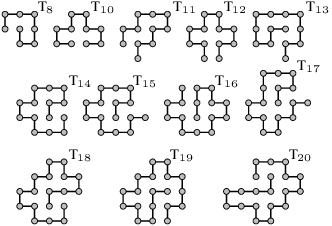}
\caption{The target structures used for the sequence optimization in Sec.~\ref{sec:qpu_results} and the structure used for verification in Appendix~\ref{sec:appC} ($\mathrm{T}_8$). The different compositions used can be found in Table~\ref{tab:qpu_problems}.
\label{fig:small_targets}}
\end{figure}

Table~\ref{tab:qpu_problems} shows the minimum $\EHP$ and the degeneracy of the solution to the
sequence optimization problem for the systems studied using pure QPU computations and
Schr\"odinger simulations, all with $N\le20$. How many of the solutions
that actually have the target structure as their unique ground state is indicated within parentheses.
This number is known from exact enumerations for these system sizes~\cite{Irback:02}. Finally,
Table~\ref{tab:qpu_problems} also shows the energy gap, $\Delta E$, between the ground state
and the first excited state of the problem Hamiltonian.  All target structures T$_{N}$ with $N\le20$
can be found in Fig.~\ref{fig:small_targets}.

\newpage

\begin{table}[H]
  \centering
  \caption{Minimum $\EHP$, the degeneracy of the solution, the energy gap $\Delta E$, and the
  chain strength $J_{cs}$ used,
  for all sequence optimization instances studied in Sec.~\ref{sec:qpu_results},
  using the target structures T$_{10}$-T$_{20}$ and different
  compositions, $\NH$. Also included is the system used for verification
  in Appendix~\ref{sec:appC} ($\mathrm{T}_8$). The number of the solutions
  that have the target structure as its unique ground state is given within parentheses.
  $\Delta E$ is the gap between the two lowest values of $E(\sv)$ [Eq.~(\ref{eq:E})], when using
  $\lambda=1.1$.
  \label{tab:qpu_problems}}
  \vspace{6pt}
  \begin{tabular}{c@{\hspace{.5cm}}c@{\hspace{.5cm}}c@{\hspace{.5cm}}c@{\hspace{.5cm}}c@{\hspace{.5cm}}c}
    \hline
    Target structure & $\NH$ & Minimum $\EHP$ & Degeneracy & $\Delta E$ &  $J_{cs}$ \\
     \hline
    T$_{8}$		& 4	& $-3$	& 1\ (1)	& 1.0	& --	\\
    \hline
    T$_{10}$	& 4	& $-4$	& 1\ (1)	& 1.1	& 2.25	\\
     \hline
    T$_{11}$	& 5	& $-4$ 	& 1\ (1)	& 1.0 &	2.25 \\

     \hline
     T$_{12}$     	& 4& $-4$ 	& 1\ (1)	& 1.1	&  2.25	\\
                    	& 6 & $-5$ 	& 1\ (1)	& 1.0	& 2.75	\\
     \hline
     T$_{13}$     	& 6 & $-4$ 	& 18\ (1)	& 0.1	& 2.75	\\
                    	& 8 & $-6$ 	& 1\ (1)	& 1.0	& 2.75	\\
     \hline
     T$_{14}$     	& 6 & $-5$ 	& 5\ (1)	& 0.1	&  3.00	\\
                    	& 8 & $-7$	& 1\ (1)	& 1.0	&  3.00	\\
     \hline
     T$_{15}$     	& 5 & $-4$ 	& 6\ (1)	& 0.1	& 3.25	\\
                    	& 6 & $-5$ 	& 5\ (1)	& 0.1	& 3.00	\\
     \hline
     T$_{16}$     	& 6 & $-6$ 	& 1\ (1)	& 0.1	& 3.00	\\
                    	& 7 & $-7$ 	& 1\ (1)	& 1.0	& 3.00	\\
                    	& 8 & $-7$	& 10\ (5)	& 0.1	& 3.25	\\
     \hline
     T$_{17}$     	& 6 & $-6$ 	& 1\ (1)	& 1.0	& 3.50	\\
                    	& 7 & $-6$	& 14\ (5)	& 0.1	& 3.50	\\
     \hline
     T$_{18}$     	& 8 & $-8$ 	& 1\ (1)	& 0.1	& 3.50	\\
                    	& 9 & $-9$ 	& 1\ (1)	& 1.0	& 3.75	\\
     \hline
     T$_{19}$     	& 8 & $-8$ 	& 1\ (1)	& 0.1	& 3.75	\\
                    	& 9 & $-9$ 	& 1\ (1)	& 1.0	& 4.00	\\
     \hline
     T$_{20}$     	& 8	& $-8$ 	& 1\ (1)	& 0.1	& 4.25	\\
                    	& 9 & $-9$ 	& 1\ (1)	& 1.0	& 4.00	\\
                    	& 11 & $-10$	& 1\ (1)	& 1.0	& 4.25	\\
    \hline
  \end{tabular}
\end{table}


\section{Numerical integration of the time-dependent Schr\"odinger equation\label{sec:appC}}

To numerically integrate the Schr\"odinger equation, we first split the time evolution
into $M$ steps with length $\epsilon$ ($t_f=M\epsilon$), yielding
\begin{equation}
\psi(t_f)=U(t_M, t_{M-1})\ldots U(t_1,t_0)\psi(0)\,,
\end{equation}
where $\psi(t)$ is the wave function corresponding to the Hamiltonian in Eq.~(\ref{eq:Ht}), $t_m=m\epsilon$ ($m=0,\ldots,M$), and
\begin{equation}
U(t_{m+1},t_m)=\mathcal{T}\exp\left[-i\int_{t_m}^{t_{m+1}}dt H(t)\right]
\end{equation}
(with $\hbar=1$). Assuming a linear $t$-dependence of $a(t)$ and $b(t)$ (Sec.~\ref{sec:methods_Schrodinger}),
a leading-order Magnus expansion~\cite{Magnus:54} yields
$\ln U(t_{m+1},t_m)\approx -i\epsilon H_m + O(\epsilon^3)$, where $H_m=H(t_m+\epsilon/2)$.
It follows that
\begin{equation}
\tilde U(t_{m+1},t_m)=\exp(-i\epsilon H_m)
\label{eq:magnus}
\end{equation}
provides a unitary, second-order accurate integrator (cubic local error). With this approximation,
the evolution from time $t_m$ to time $t_{m+1}$ is governed by the
constant Hamiltonian $H_m$. Still, due to the Hilbert space dimensionality ($2^N$),
the numerical implementation requires care. To this end, we replace
$\tilde U(t_{m+1},t_m)$ [Eq.~(\ref{eq:magnus})]
by the Crank-Nicolson integrator~\cite{Crank:47}
\begin{equation}
\doubletilde{U} (t_{m+1},t_m)=(\mathbf{T}^{\dag})^{-1}\mathbf{T} \quad\mathrm{where}\quad \mathbf{T}=1-\frac{i\epsilon}{2}H_m\,,
\label{eq:CN}
\end{equation}
which is implicit but still feasible, thanks to the sparseness of $H_m$.
Like $\tilde U(t_{m+1},t_m)$, $\doubletilde{U}(t_{m+1},t_m)$ is unitary and second-order accurate.
The sparseness of $H_m$ can be easily exploited by rewriting the relation
$\psi(t_{m+1})=\doubletilde{U}(t_{m+1},t_m)\psi(t_m)$ as
\begin{equation}
\mathbf{A}\mathbf{v}=\mathbf{u}\quad\text{with}\quad \mathbf{A}=\mathbf{T}\mathbf{T}^\dag\ \text{and}\ \mathbf{u}=\mathbf{T}^2\psi(t_m)\,,
\label{eq:CG}
\end{equation}
and solving this linear system of equations for $\mathbf{v}=\psi(t_{m+1})$ by the conjugate gradient method~\cite{Hestenes:52}.
In Eq.~(\ref{eq:CG}), both sides of the equation were multiplied by $\mathbf{T}$, in order to have
a Hermitian and positive definite matrix $\mathbf{A}$, as required by the conjugate gradient method.
\begin{figure}[t]
\centering
  \includegraphics[width=7cm,angle=0]{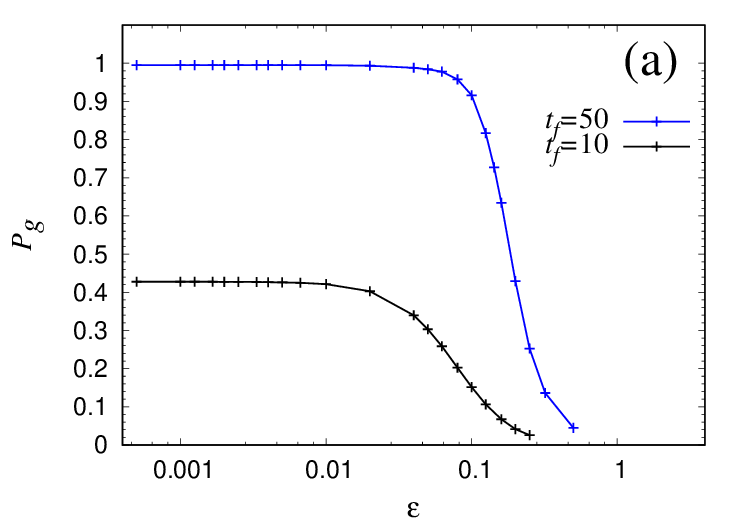}\quad
  \includegraphics[width=7cm,angle=0]{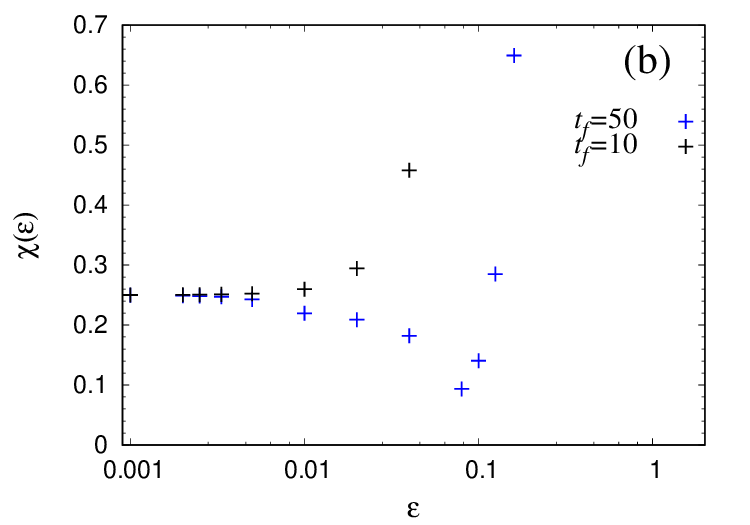}\quad
\caption{Step size dependencies when using Eq.~(\ref{eq:CN})
to integrate the Schr\"odinger equation of an eight-qubit system ($\mathrm{T}_8$ [Fig.~\ref{fig:small_targets}] with $\NH=4$ [Table \ref{tab:qpu_problems}]) for two
choices of the annealing time, $t_f=10$ and $t_f=50$. (a) The probability of finding the
final system in the ground state of the problem Hamiltonian $\HP$, $P_g$,
plotted against the step size $\epsilon$. The lines are drawn to guide the eye. (b) The $\epsilon$-dependence of the finite-difference
ratio $\chi(\epsilon)=[P_g(\epsilon)-P_g(\epsilon/2)]/[P_g(2\epsilon)-P_g(\epsilon)]$,
which approaches $2^{-k}$ for small $\epsilon$ if the integrator is $k$th-order accurate.
The data for $\chi(\epsilon)$ fall close to 0.25 for small $\epsilon$, as expected
for the second-order integrator in Eq.~(\ref{eq:CN}).
\label{fig:eps}}
\end{figure}
We implemented this algorithm, based on the Crank-Nicolson and conjugate-gradient methods, into a C$++$ program.
Figure~\ref{fig:eps} shows results from a test of the program on an eight-qubit system, using two choices of annealing
time, $t_f=10$ and $t_f=50$, and different step sizes $\epsilon$. From panel (a) it can be seen that the value $t_f=50$
if sufficiently large for the ground-state probability at $t=t_f$, $P_g$, to be close to 1, which is not case for $t_f=10$.
Panel (b) shows data for a finite-difference ratio, $\chi(\epsilon)$, which confirm that, for fixed $t_f$, $P_g$ displays
the expected quadratic dependence on $\epsilon$ for small $\epsilon$.


\section{Pure QPU hit rates with and without chain breaks\label{sec:appD}}

\begin{figure}[t]
\centering
   \includegraphics[width=8cm]{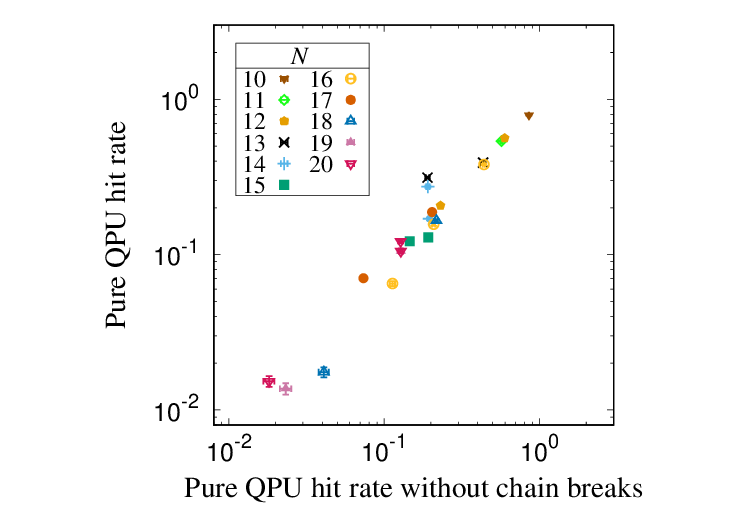}
\caption{Scatter plot comparing pure QPU hit rates computed using all data [Fig.~\ref{fig:hit_qpu}(a)]
to those obtained after filtering out all annealing cycles in which any chain break occurred.}
\label{fig:chainbreaks}
\end{figure}

In our pure QPU computations, using the \texttt{DWaveCliqueSampler},
each logical qubit is represented by a chain of two or three physical qubits.
It may happen that such chains break. The pure QPU hit rates shown in
Fig.~\ref{fig:hit_qpu}(a) represent averages over all generated annealing
cycles, irrespective of whether or not all chains were intact. Fig.~\ref{fig:chainbreaks}
compares these hit rates with those obtained when omitting from the analysis
all annealing cycles in which any chain break occurred (23\,\% of the data).
The results obtained with and without this filter are similar (Pearson correlation coefficient 0.94).

\begin{acknowledgments}
This work was in part supported by the Swedish
Research Council (Grant no.~621-2018-04976).
We gratefully acknowledge the J\"ulich Supercomputing Centre
(https://www.fz-juelich.de/ias/jsc) for supporting this project by providing computing time
on the D-Wave Advantage\texttrademark{} System JUPSI through the J\"ulich UNified
Infrastructure for Quantum computing (JUNIQ). We acknowledge helpful conversations with Ken Robbins from D-Wave,
and have also benefitted from discussions with Göran Johansson, Hanna Linn and Laura Garc\'{\i}a-\'Alvarez
at the Wallenberg Centre for Quantum Technology at Chalmers University of Technology.
\end{acknowledgments}

%

\end{document}